\documentclass{rmf-d}
\usepackage{nopageno,rmfbib,multicol,times,epsf,amsmath,amssymb,cite}
\usepackage[latin1]{inputenc}
\usepackage{graphics}
\usepackage{float}

\usepackage{tikz}

\usepackage[compat=1.1.0]{tikz-feynman}

\def\rmfcornisa{Research OR Education of Physics \hfill\rmf\ {\bf ??} (*?*) ???--???
\hfill MES? A\~NO?}

\def\rmfcintilla{{\it Rev.\ Mex.\ Fis.\/} {\bf ??} (*?*) (????) ???--???}
\begin{document}
\title{   Study of bound states in a thermal gas using the S-matrix formalism
\vspace{-6pt}}
\author{ Subhasis Samanta  }
\address{ \textit{Institute of Physics, Jan-Kochanowski University, }\\\textit{ul. Uniwersytecka 7, 25-406 Kielce, Poland.}    }
\author{ }
\address{ }
\author{ }
\address{ }
\author{ }
\address{ }
\author{ }
\address{ }
\maketitle
\recibido{day month year}{day month year
\vspace{-12pt}}
\begin{abstract}
\vspace{1em} 

We have studied the formation of bound states in a thermal gas in the context of quantum field theory
(QFT). We have considered a scalar QFT with $\varphi^4$ interaction, where $\varphi$ is a scalar particle with mass $m$. We have observed the formation of a bound state of $\varphi$-$\varphi$ type when the coupling constant is negative and its modulus is larger than a certain critical value. 
We have calculated the contribution of the bound state to the pressure of the thermal gas of the system by using the S-matrix formalism. Our analysis is based on a unitarized one-loop resumed approach in which the theory is finite and well defined for each value of the coupling constant. We have observed that the total pressure
as a function of the coupling constant is continuous also at the critical coupling: the jump in pressure due to the sudden appearance of the bound state is exactly cancelled by an
analogous jump (but with opposite sign) of the interaction contribution to the pressure.

\vspace{1em}
\end{abstract}
\keys{Bound states, Particle interactions, Scattering phase shifts, Scattering amplitudes  \vspace{-4pt}}
\begin{multicols}{2}


The production of hadronic bound states such as deuteron ($d$), nucleus of helium-3 ($^{3}$He), tritium ($^{3}\text{H}$), helium-4 ($^{4}\text{He}$), hypertritium
($_{\Lambda}^{3}$H) and their antiparticles have attracted a lot of interest because their binding energies are typically much smaller than the temperature realized in high energy collisions \cite{ Abelev:2010rv,Agakishiev:2011ib,Adam:2015vda,Adam:2019phl,Acharya:2019xmu,Acharya:2020sfy}. It is quite puzzling how these objects can form in such a hot environment.  

We intend to investigate these questions in the context of Quantum Field Theory (QFT). The temperature dependence is
incorporated by using the S-matrix (or phase-shift) formalism. This proceeding is based on Ref.  \cite{Samanta:2020pez}.

We use a scalar $\lambda \varphi^4$ \cite{Samanta:2020pez} 
type interaction where $\lambda$ is the dimensionless coupling constant. At the tree level, this interaction 
corresponds to a delta-potential in the non-relativistic limit \cite{Beg:1984yh}.
The Lagrangian of the study reads

\begin{equation}
\mathcal{L}=\frac{1}{2}\left(  \partial_{\mu}\varphi\right)  ^{2}-\frac{1}%
{2}m^{2}\varphi^{2}-\frac{\lambda}{4!}\varphi^{4}, \label{eq:L}%
\end{equation}
where the first two terms describe a free scalar particle with mass $m$ and the last term corresponds to the interaction. 
In the centre of mass frame, the differential cross-section can be written as \cite{Peskin:1995ev}
\begin{equation}
\frac{d\sigma}{d\Omega}=\frac{|A(s,t,u)|^{2}}{64\pi^{2}s}~,
\end{equation}
where $A(s,t,u)$ is the scattering amplitude as evaluated through Feynman
diagrams, and $s,t$ and $u$ are usual Mandelstam variables.
The sum of these three variables is $s+t+u=4m^{2}$. In terms of partial waves, the scattering amplitude can be expressed in terms of
$s$ and scattering angle $\theta$ as \cite{Messiah}:%

\begin{equation}
A(s,t,u)=A(s,\theta)=\sum_{l=0}^{\infty}(2l+1)A_{l}(s)P_{l}(\cos
\theta)\text{{ ,}}%
\end{equation}
where $P_{l}(\xi)$ with $\xi=\cos\theta$ are the Legendre polynomials.
In general, the $l$-th wave partial amplitude is given by
\begin{equation}
 A_{l}(s)=\frac{1}{2}\int_{-1}^{+1}d\xi A(s,\theta)P_{l}(\xi).
\end{equation}

In the particular case of our Lagrangian of Eq. (\ref{eq:L}), the tree-level 
scattering amplitude $A(s,t,u)$ takes the form:
\begin{equation}
iA(s,t,u)=i(-\lambda)\Rightarrow A(s,t,u)=A(s,\theta)=-\lambda.
\end{equation}
The scattering amplitude $A<0$ for $\lambda>0$. This implies that the tree-level interaction is repulsive. On the other hand for $\lambda<0$, $A>0,$ which corresponds to an attractive interaction.

The $s$-wave ($l=0$) amplitude takes the form:
\begin{equation}
A_{0}(s)=\frac{1}{2}\int_{-1}^{+1}d\xi A(s,\theta)=A(s,\theta)=-\lambda\text{
,}%
\end{equation}
while for all other partial waves $A_{l=1,2,...}(s)=0$ (this
holds true also when unitarizing the theory within the adopted resummation scheme, see below). Therefore, the total cross-section is given by
\begin{equation}
\sigma(s)=\frac{1}{2}2\pi\frac{1}{64\pi^{2}s}\sum_{l=0}^{\infty}%
2(2l+1)\left\vert A_{l}(s)\right\vert ^{2}=\frac{\lambda^2}{32\pi%
s}. \label{eq:sigma}
\end{equation}
At the threshold:%
\begin{equation}
\sigma(s_{th}=4m^{2})=\frac{1}{2}2\pi\frac{1}{64\pi^{2}s}2 \lambda^{2}=8\pi\left\vert a_{0}^{\text{SL}}\right\vert
^{2},
\end{equation}
where $a_{0}^{\text{SL}}$ is the tree-level s-wave scattering length:%
\begin{equation}
a_{0}^{\text{SL}}=\frac{1}{2}\frac{A_{0}(s=4m^{2})}{8\pi\sqrt{4m^{2}}}%
=\frac{1}{2}\frac{-\lambda}{16\pi m}\text{ .}%
\end{equation}
The factor $1/2$ in the previous equation is due to the identical particles.

\subsection*{Unitarization}

\label{sec:Unitarization} 
\begin{figure}[H]
 \centering
   \includegraphics[width=0.15\textwidth]{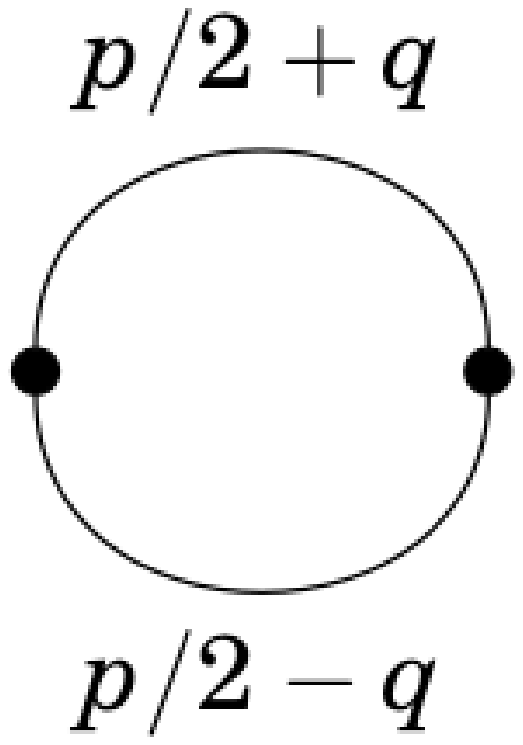}
   \caption{The loop $\Sigma(p^2)$ (Eq. \ref{eq:loop1}) used for the unitarization.}
   \label{fig:loop}
 \end{figure}

Let us now discuss the method of unitarization implemented in this work. For this, we introduce the two-particle loop ($\Sigma(p^2)$) of the field $\varphi$ \cite{Giacosa:2007bn} as shown in Fig. \ref{fig:loop}.

 The loop function $\Sigma(p^2)$ can be written as
 \begin{equation}
 \begin{split}
     & \Sigma(p^2) = -i \int \frac{d^4q}{(2\pi)^4} \\& \frac{1}{[(p/2-q)^2-m^2+i \varepsilon][(p/2+q)^2-m^2+i \varepsilon]},
 \end{split}    
 \label{eq:loop1}
 \end{equation}

where $p = p_1+p_2 = p_3+p_4$ ($p_{1},p_{2}$ are four-momenta of ingoing particles and $p_{3},p_{4}$ are the same for outgoing particles).

 The imaginary part of $\Sigma(s)$ above the threshold is required to be
\begin{equation}
\label{eq:ImSigma}
I(s)=\operatorname{Im}\Sigma(s)=\frac{1}{2}\frac{\sqrt{\frac{s}{4}-m^{2}}%
}{8\pi\sqrt{s}}\text{for }\sqrt{s}>2m . 
\end{equation}
This requirement is due to the optical theorem. The above equation is considered valid up to arbitrary values of the variable $s$ since we have not put any cutoff.
The function $\Sigma(s)$ for complex values of the variable $s$ reads
\begin{equation}
\Sigma(s)=\frac{1}{\pi}\int_{4m^{2}}^{\infty}ds^{\prime}\frac{I(s^{\prime}%
)}{s^{\prime}-s-i\epsilon}-C\text{ ,}
\end{equation}
where $C$ is a subtraction that guarantees convergence of the loop function. Here, we make the choice
$\Sigma(s\rightarrow0)=0,$ hence
\begin{equation}
C=\frac{1}{\pi}\int_{4m^{2}}^{\infty}ds^{\prime}\frac{I(s^{\prime})}%
{s^{\prime}}.
\end{equation}
Finally, the
loop reads:%

\begin{equation}
\Sigma(s)=\frac{1}{2}\frac{1}{16\pi}\left(  -\frac{1}{\pi}\sqrt{1-\frac
{4m^{2}}{s+i\epsilon}}\ln\frac{\sqrt{1-\frac{4m^{2}}{s+i\epsilon}}+1}%
{\sqrt{1-\frac{4m^{2}}{s+i\epsilon}}-1}\right)  +\frac{1}{16\pi^{2}}\text{.}
\label{eq:loop}%
\end{equation}

\begin{figure}[H]
 \centering
 \includegraphics[width=0.3\textwidth]{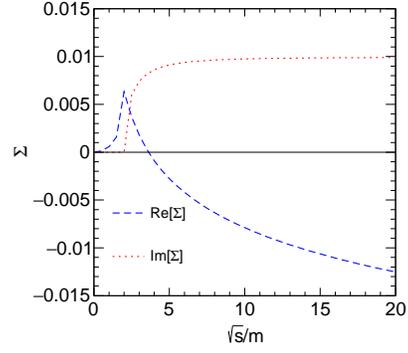}
 \caption{Real and imaginary parts of $\Sigma$ as a function of $\sqrt{s}/m$.}
 \label{fig:loop_re_im}
\end{figure}
Figure \ref{fig:loop_re_im} shows real and imaginary parts of $\Sigma$ as a function of $\sqrt{s}/m$. The real $\Sigma(s)$ is zero at $\sqrt{s} = 0$ and reaches a maxumum at the threshold $\sqrt{s} = 2m$ then decreases monotonically and becomes negative for $\sqrt{s}/m$ large enough. The imaginary part of $\Sigma$ is infinitesimally small below the threshold, while above threshold it increases according to Eq. \ref{eq:ImSigma}.

The unitarized amplitudes in the
$k$-channel is calculated using one loop resumed approach as follows:
\begin{equation}
A_{k}^{U}(s)=\left[  A_{k}^{-1}(s)-\Sigma(s)\right]  ^{-1}.
\end{equation}
Only the unitarized s-wave amplitude is
nonzero and takes the form:
\begin{equation}
A_{0}^{U}(s)=\left[  A_{0}^{-1}(s)-\Sigma(s)\right]  ^{-1}=\frac{-\lambda
}{1+\lambda\Sigma(s)}\text{ .}%
\end{equation}%

 
\subsection*{Bound state}
\label{sec:BoundStateFormation}
If $\lambda$ is negative the two scalar particles attract each other. When the attraction is large enough a bound state emerges. The unitarized amplitude diverges when a bound state is formed. So the bound state equation reads:
\begin{equation}
A_{0}^{U}(s)^{-1}=\left[  -\lambda^{-1}-\Sigma(s)\right]  =0\text{ .}
\label{bseq}%
\end{equation}
Since $\Sigma(s)$ is solely real for $s<4m^{2}$ and has a maximum at threshold with $\Sigma(s=4m^{2})=\frac{1}{16\pi^{2}}$, it turns out
that a bound state is present if
\begin{equation}
\lambda\leq \lambda_{c}=-16\pi^{2}.%
\end{equation}

\begin{figure}[H]
 \includegraphics[width=0.3\textwidth]{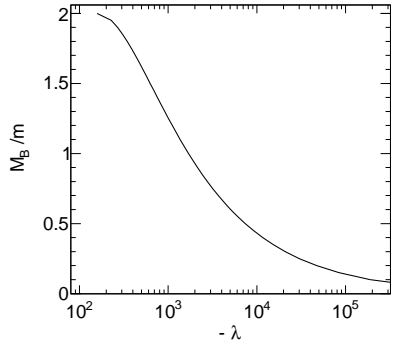}
 \caption{$\lambda$ dependence of bound state mass.}
 \label{fig:mb_vs_lambda}
\end{figure}

The mass ratio $M_{B}/m$ as a function of $\lambda$, plotted in Fig.
\ref{fig:mb_vs_lambda}, fulfills the conditions:
\begin{align}
M_{B}(\lambda &  =\lambda_{c})=2m\text{ ,}\\
M_{B}(\lambda &  \rightarrow-\infty)=0\text{ ,}%
\end{align}
the latter being a consequence of the employed unitarization.
For identical particles, the s wave phase shift can be calculated from the following definition:

\begin{equation}
\frac{e^{2i\delta_{0}^{U}(s)}-1}{2i}=\frac{1}{2}\cdot\frac{k}{8\pi\sqrt{s}%
}A_{0}^{U}(s)\text{ .} \label{delta0U}%
\end{equation}
where $k=\sqrt{\frac{s}{4}-m^{2}}$ is the modulus of the three-momentum of one of the ingoing (or outgoing) particles.


 \begin{figure}[H]
  \centering
  \includegraphics[width=0.3\textwidth]{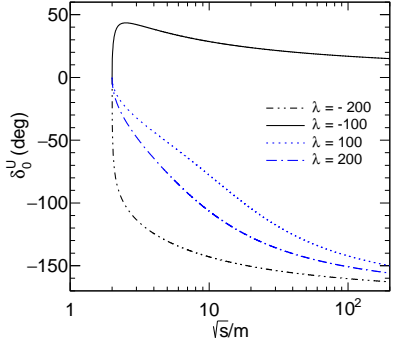}
  \caption{Uniterized s-wave phase shifts as a function of $\sqrt{s}/m$ for different $\lambda$ values.}
  \label{fig:ps}
 \end{figure}

Figure \ref{fig:ps} shows the unitarized s-wave phase shifts as a function of $\sqrt{s}/m$ for different $\lambda$ values. We observe that for all the cases phase shifts follow the Levinson theorem \cite{Hartle:1965nj}
\begin{equation}
n_{\text{poles-below-threshold}}=\frac{1}{\pi}\left(  \delta_{0}%
^{U}(s\rightarrow\infty^{2})-\delta_{0}^{U}(s=4m^{2})\right)  \text{ .}%
\end{equation}





Now we will study the system at finite temperature. The non-interacting part of the pressure for a gas of particles with mass $m$ reads:%
 
\begin{equation}
P_{\varphi\text{,free}}=-T\int_{k}\ln\left[  1-e^{-\sqrt{k^{2}+m^{2}}/T%
}\right]  \text{,}%
\end{equation}
where $\int_k \equiv \int d^3k/(2\pi)^3$.

In the $S$-matrix formalism \cite{Dashen:1969ep,Venugopalan:1992hy,Broniowski:2015oha,Lo:2017ldt,Lo:2017sde,Lo:2017lym,Dash:2018mep,Lo:2019who}, the interacting part of the pressure is related to the derivative of the
phase shift with respect to the energy by the following relation:%
\begin{equation}
\begin{split}
 P_{\varphi\varphi\text{-int}}=-T\int_{2m}^{\infty}dx\frac{1}{\pi}\frac
{d\delta_{0}(s=x^{2})}{dx} \int_{k}\ln\left[  1-e^{-\sqrt{k^{2}+x^{2}}/T%
}\right]  \text{ ,}%
\end{split}
\end{equation}
where $x=\sqrt{s}$.


The crucial question of the present work is how to include the effect of the emergent bound state $B$ in the
thermodynamics. The bound state contribution to the pressure can be writen as:

\begin{equation}
P_{B}\overset{\text{ }}{=}-\theta(\lambda_{c}-\lambda)T\int_{k}\ln\left[
1-e^{-\sqrt{k^{2}+M_{B}^{2}}/T}\right]  \label{pb},%
\end{equation}
where the theta function takes into account that for $\lambda>\lambda_{c}$
there is no bound state $B.$

 \begin{figure}[H]
  \includegraphics[width=0.3\textwidth]{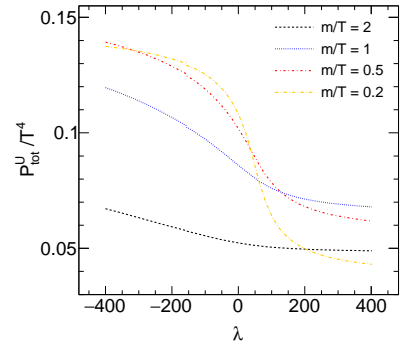}
  \caption{Variation of normalized total unitarized pressure with $\lambda$. }
  \label{fig:Ptot}
 \end{figure}
  
\begin{figure}[H]
\centering
\includegraphics[width=0.3\textwidth]{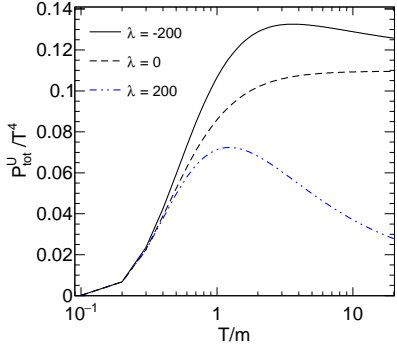}
\caption{Temperature dependence of the total normalized pressure.}%
\label{fig:P_tot_vs_T_uniterized}
\end{figure}

\begin{figure}[H]
\centering
\includegraphics[width=0.3\textwidth]{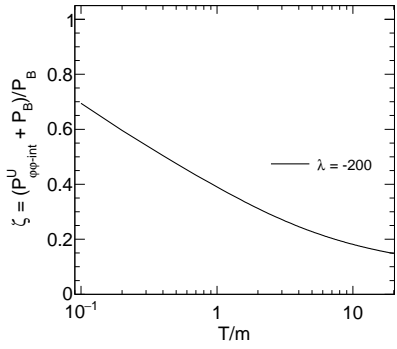}
\caption{Variation of $\zeta$ with $T/m$ for $\lambda = -200$.}%
\label{fig:fraction_of_bs}
\end{figure}

The full unitarized pressure looks like
\begin{equation}
P_{tot}^{U}=P_{B}+P_{\varphi\text{,free}}+P_{\varphi\varphi\text{-int}}%
^{U} .%
\end{equation}
Remarkably, $P_{tot}^{U}$, shown in Fig. \ref{fig:Ptot}, turns out to be a continuous function of
$\lambda$, even if $P_{B}$ and $P_{\varphi \varphi\text{-int}}^{U}$ are not continuous at $\lambda=\lambda_{c}$. $P_{tot}^{U}$ as a function of $T/m$ is shown in Fig. \ref{fig:P_tot_vs_T_uniterized}. Finally, for the illustrative value $\lambda =-200$, in Fig.\ref{fig:fraction_of_bs} we show
the temperature dependence of the following quantity:
\begin{equation}
\zeta(T,\lambda)=\frac{P_{\varphi\varphi\text{-int}}^{U}+P_{B}}{P_{B}}.%
\end{equation}
The Fig. \ref{fig:fraction_of_bs} indicates that the total interacting contribution to the pressure including both the bound state and the $\varphi$-$\varphi$  interaction
above threshold $\zeta P_B$ with $0 < \zeta <1$.
   
\section*{Summary}

In this work, we have investigated the formation of bound states in a thermal gas in the context of selected scalar QFT that contain four-leg vertices.
We have employed a unitarized one-loop resummed approach for which the theory is finite and well defined for each value of the coupling constant $\lambda$. Moreover, a bound state is formed when attraction is large enough ($\lambda \le \lambda_c$).
The s-wave scattering phase shift has been calculated using the partial wave decomposition of two-body scattering. Further, the temperature dependence is included using the S-matrix or phase-shift  approach, according to which the density of the state is proportional to the derivative of the phase shift with respect to $\sqrt{s}$. Quite remarkably, the total
pressure as a function of $\lambda$ is continuous. The jump in pressure generated by the abrupt appearance of the bound state is exactly cancelled by
an analogous jump in the opposite sign due to the phase-shift contribution to the pressure. Further, the contribution of the bound state to the total pressure is partially cancelled by $\varphi$-$\varphi$  interaction
above the threshold.

In Ref. \cite{Samanta:2021vgt} we have extended this work by considering more complex QFTs
with three-leg interaction
vertices, $g \varphi^3$ and $g S\varphi^2$ ($S$ being also a scalar field), which constitute the relativistic counterpart
of the Yukawa potential. In these interactions, besides the s-channel, also the t-channel and u-channel Feynman diagrams contribute.
In both types of interactions a $\varphi \varphi$ bound state
is formed when the coupling $g$ is greater than a certain critical value. Similar to the $\varphi^4$ interaction, a partial cancellation between the negative contribution of the interaction part of the pressure with the positive one of the bound state pressure occurs. Further, the total pressure is continuous as a function of the coupling $g$.

\section*{Acknowledgement}
The author acknowledges financial support from Polish National Agency for Academic Exchange (NAWA) through the Ulam Scholarship with the agreement no:
PPN/ULM/2019/1/00093/U/00001.
Author thanks Francesco Giacosa for useful discussions.

\end{multicols}
\medline
\begin{multicols}{2}

\end{multicols}
\end{document}